\documentstyle[preprint,aps,graphicx]{revtex}
\tightenlines
\begin{document}
\newcommand{\bref}[1]{eq.~(\ref{#1})}
\newcommand{\be}{\begin{equation}}
\newcommand{\en}{\end{equation}}
\newcommand{\bs}{$\backslash$}
\newcommand{\us}{$\_$}

\title{Optical detection of magnetic data using magnetooptic indicators 
with in-plane magnetization}
\author{L.E. Helseth, E.I. Il'yashenko, M. Baziljevich and T.H. Johansen}
\address{Department of Physics, University of Oslo, P.O. Box 1048 Blindern,
N-0316 Oslo, Norway\\
email: l.e.helseth@fys.uio.no
}

\maketitle
\begin{abstract}
We investigate detection of magnetic data using 
magnetooptical indicators with in-plane magnetization. A simple model for the
magnetooptical detection system is presented. We find that the signal to noise ratio changes strongly with the bit size, the 
polarization noise and the distance between the magnetic carrier and the indicator. In
particular, it is found that within our model a signal to noise ratio of 30 is
possible for a bit size of 200 nm. We also estimate the resolution of the 
system, and find that a spot size of $\sim 200$ nm can be obtained using a 
suitably designed solid immersion lens. Finally, we 
discuss detection of several magnetic tracks simultaneously.        
\end{abstract}

\narrowtext

\newpage

\section{Introduction}
Magnetic data storage and readout has long been an active field of research due
to the ever increasing demands for faster and better access to stored data. The 
areal density for information storage on magnetic media has increased at an 
astonishing rate over the last three decades. From 1970 - 1990 
the density increased at a rate of $30\%$ per year. In 1992 IBM introduced the
magnetoresistive read-head, and the growth rate increased to $60\%$ per year.
A very successful development followed in 1998 when IBM announced giant
magnetoresistive spin-valves, and thus boosted the growth rate to 
$100\%$ per year\cite{Grady}. As the bit density gets higher, it becomes increasingly 
difficult and costly to design and construct media, read-head, write head, 
tracking system and error correction electronics. 

Many attempts have been directed towards reading out magnetic data by optical
systems, since optical readout systems have many interesting features not
possessed by magnetic readout heads\cite{Mansuripur,Gambino,Hatakeyama,Tokumaru}. 
However, in most optical systems the resolution is limited by the classical diffraction limit 
$\sim \lambda/2NA$, where $\lambda$ is the wavelength of light, and NA is the 
numerical aperture.  
A possible way to circumvent this problem was introduced by japanese 
researchers in the 1980's \cite{Hatakeyama,Tokumaru}. In 1986, Hatakeyama
et al. \cite{Hatakeyama}
reported that when a thin magnetic film is positioned 
normally to the recording medium surface, i.e., similar to a single-pole 
magnetoresistive head, a higher linear resolution may be expected. To 
demonstrate the principle, they sputtered a Co-Zr thin film on glass, and added a
nonmagnetic mirror. By utilizing the longitudinal Kerr effect, they were 
able to demonstrate a resolution of $\sim 1$ $\mu m$, but claimed that the
resolution prospects are much better than this.  Soon after, Tokumaru and 
Nomura demonstrated a different system with approximately the same resolution 
prospects\cite{Tokumaru}. Their system consisted of a high permeability 
thin film placed in perpendicular contact with the recording medium, thus
guiding the magnetic signal to a stripe-domain Bi-substituted iron garnet
film, which responded to the field variations by changing the width of its
stripes.  
In spite of these two very interesting proposals, no further development has
taken place on such hybrid optical and magnetical storage systems. 

In this paper we analyze a different approach, based on
ref.\cite{Il'yashenko}, for optical detection of
magnetically stored data and other weak magnetic flux sources. The basic system, 
shown in Fig. \ref{f1}, is in principle a polarization microscope which 
detects the magnetization distribution of a magnetooptic indicator\cite{Antonov,Jevenko,Belyaeva,Johansen}. 
Like most other optical systems, our system is also limited by the classical diffraction limit,
but it will be shown that by utilizing a suitable solid immersion lens it is 
possible to obtain a spot size of $\sim 200$ nm. 
If the indicator has virtually no domains and is sufficiently sensitive to 
external fields, it will redistribute its magnetization within less than a
nanosecond, and follow the field\cite{Freeman}. Here we present a simple model, which describes the magnetization in 
the indicator and the magnetooptical detection system. We evaluate
the signal to noise ratio as well as the resolution and the focal depth. 
Finally, we discuss detection of several magnetic tracks simultaneously. 

\section{Readout principle}
Today there exists different methods for magnetooptical readout; 
the single ended detection system of Fig. \ref{f1} and the
differential detection system of Fig. \ref{f2}. The differential detection system 
is a key feature to the success of modern magnetooptical storage devices, mainly 
due to its low noise and superior signal. 
The light source is here a polarized semiconductor laser diode operating at constant 
output power (in practice the laser's drive current is modulated at several 
hundred MHz to reduce noise from e.g. mode hopping and mode competition). The 
light beam from the laser is then collimated and passed through a polarizer to 
ensure that all the light is x-polarized.  This light is then focused 
by the combined action of an objective and a spherical solid immersion lens (SIL). 
The SIL is placed in very close contact with a magnetooptic indicator. To 
obtain the best possible contact, the indicator should be deposited directly onto the 
surface of the SIL. To reflect the light (both x and y-polarized), a mirror is
deposited onto the indicator. 
This mirror should also function as a protection against wear from the magnetic carrier medium. 
If this is not sufficient, an additional protective layer (e.g. $Ti_{3}N_{4}$
or equivalent) can be applied. The magnetic carrier is moved while staying in close contact with the 
mirror/protective layer. Usually an airgap is formed between the two, and the sum of the 
airgap and the total thickness of the mirror/protective layer will be called the flying height 
of the carrier, since this corresponds to the actual distance to the magnetic
sensor.       

The indicator will redistribute its magnetization according to the magnetic source. 
Furthermore, since the indicator is supposed to have 
large magnetooptic effect, it will turn some of the incident x - polarized light 
into y - polarized light. In the return path, the partially polarizing beamsplitter 
(PPB) provides the detection module with a fraction $\gamma$ of the x-component and 
the entire y-component. The x-component, although without significant information, 
is needed for proper extraction of data\cite{Mansuripur}.
A polarizing beam splitter (PB), with its axes at $45^{\circ}$ to the x and y-polarization, 
splits the beam in two components. Together with detectors 1 and 2, the PB comprise a 
balanced differential detection module. The detectors are low noise photodiodes,
and their signals form the input of a differential amplifier.

\section{Magnetooptic response}
In order to serve as an indicator, the film must in addition to having a small
anisotropy field, also have a large magnetooptic effect, as shown in 
Fig. \ref{f3}, and we will here only consider the Faraday rotation.
Due to this effect, secondary light beams are generated on the way down and also after
reflection from the mirror. 
The combined action of the mirror and the indicator is represented by the 
effective reflection matrix
\begin{displaymath}
\mathbf{R} = 
\left[ \begin{array}{ccc} 
r_{x} &r_{xy} \\
r_{yx} & r_{y} \\
\end{array} \right] \,\,\,  .
\end{displaymath}
If we assume that multiple reflections 
and absorption can be neglected, and that the light is 
normally incident on the indicator, one has $r_{x}=r_{y}=1$ and\cite{Hubert1} 
\begin{equation}
r_{yx} =-r_{xy} = \frac{2V}{M_{s}}\int _{0}^{t} M_{z} dz \,\,\, . 
\label{r}
\end{equation}  
Here $V$ is a constant characteristic of the indicator, $t$ the thickness and 
$M_{s}$ is the saturation magnetization. The
z-component of the magnetization is expressed as\cite{Shamonin}
\begin{equation}
M_{z} =  M_{s} \frac{H_{z}}{H_{a}}, \,\,\,\,\,\, 
H_{a} = M_{s} -\frac{2K_{u}}{\mu _{0} M_{s}} \,\,\, , 
\end{equation}
where $H_{a}$ is the anisotropy field and $K_{u}$ is the uniaxial anisotropy 
constant. 
For a Kerr film similar expressions for the reflection coefficients 
exists\cite{Hubert1}. 

For data stored on tapes and hard discs
the magnetic field can be found rather accurately by applying the 
Landau-Lifshitz-Gilbert equations\cite{Grady,Gambino}. Such detailed 
calculations require extensive computational efforts, and will not be 
considered here. Instead we will follow Hatakeyama et al.\cite{Hatakeyama},
who assumed that the field from a source of size b has a z-component given 
by
\begin{equation}
H_{z}=M\exp\left( -\pi \frac{z+d}{b} \right) \,\,\, .
\label{Hz}
\end{equation}   
The remanent magnetization of the recording medium, $M$, will in general
depend on the bit size and the configuration of the neighbour bits. However, 
we will take it to be constant with $M=100$ kA/m. We also neglect the
presence of an in-plane magnetic field, since the main effect of such a 
component is to increase the effective anisotropy field, and thereby reducing 
the sensitivity. The eq. (\ref{r}) and eq. (\ref{Hz}) give then the 
following magnetooptic coefficient 
\begin{equation}
r_{yx} = \frac{2VMb}{\pi H_{a}} \exp\left(-\frac{\pi d}{b} \right) \left[ 1-
\exp \left( -\frac{\pi}{b} t \right) \right] \,\,\, .
\label{magref}
\end{equation}  

We define the polar Faraday rotation as
\begin{equation}
\Theta = \frac{r_{yx}}{r_{x}} \,\,\, .
\label{far}
\end{equation}

There are many materials which may function as indicators. In
particular, bismuth-substituted ferrite garnet films with in-plane magnetization 
have been shown to have very little domain activity as well as excellent
sensitivity to external fields\cite{Antonov,Jevenko,Johansen,Zvezdin,Helseth}. 
Fig. \ref{f4} shows experimental values for the Faraday rotation (at 
$\lambda =633$ nm) 
as a function of a uniform magnetic field ($H_{z}$) for two different garnet 
films. From the figure we find that $V\sim 1$ $\mu m ^{-1}$ and
$H_{a} \approx 60$ $kA/m$. By shifting the wavelength towards $\lambda =510$ nm
and increasing the bismuth content it is possible to increase V by a factor 10, although at the cost of increased 
absorption\cite{Helseth}. 

Shown in Fig. \ref{f5} is the Faraday rotation
obtained from eq. (\ref{magref}) and eq. (\ref{far}) as a 
function of bit size $b$ for two film thicknesses and d=100 nm. One sees that 
for detection of bits smaller than $0.5$ $\mu m$ one gains very little by using garnets thicker
than $0.5$ $\mu m$. Fig. \ref{f6} shows the calculated Faraday rotation as function of bit size for
different flying heights $d$ when $t=0.5$ $\mu m$. Note that the signal is
strongly dependent on the flying height. We find that flying heights of less
than 100 nm are required to obtain 0.1 degree rotation when b = 200 nm.

\section{Signal to noise ratio}
To calculate the signal to noise ratio, we will use the model considered in 
the previous section. Let us assume that the incident 
beam consists of x - polarized light only, and has
intensity $I_{0}=|E_{0}|^{2}$.
The light is first transmitted through the isotropic GGG substrate,
and then reach the magnetooptic layer, where some fraction is transformed 
into y-polarized light. After reflection the rays are transmitted once more through the substrate before
they reach the analyzer. Then the normalized electric field components just before the
analyzer can be written as $E_{x} = r_{x} = 1$ and $E_{y}=r_{yx}$ .
Here we have neglected all differences in transmission coefficients for x and
y-polarized light. 

For various reasons, there are fluctuations in
the effective reflectivity of the s and p-component of the 
polarization. First, the creation of domains and the redistribution of magnetization will 
vary rapidly with the magnetic signal, and this may give rise to a  
y-polarized noise component. Second, the noise from the magnetic carrier may 
have its own noise which is transferred to the optical signal in the form of y-polarized noise. 
Third, there are residual amounts of vibration and defocus errors that vary with time and, 
therefore, cause the x-polarized reflectivity to fluctuate. However, this class of noise is 
rather common to most readout devices (depend on mechanical design), and will not be considered here.
We will follow the procedure given by Mansuripur in dealing with the noisy signal\cite{Mansuripur}.
In this approach, we write the x and y-component of the reflectivity as
\begin{equation}
r_{x} (t) =1+\delta _{x} (t) \,\, , \,\,\,  |\delta _{x}| \ll 1\,\,\, ,
\end{equation}
and
\begin{equation}
r_{y} (t) =1+\delta _{y} (t) \,\, , \,\,\,  |\delta _{y}| \ll 1\,\,\, .
\end{equation}

There is also a conversion from x to y-polarized light not caused by
the magnetooptic effect, but rather by random scattering events. This can be
expressed as
\begin{equation}
r_{xy} (t) =\delta _{xy} (t) \,\, , \,\,\,  |\delta _{xy}| \ll 1\,\,\, ,
\end{equation}
We do not expect this component to be large, since the optical system is at rest, 
thus reducing all random scattering events to a minimum.  
We will ignore all correlations between $\delta _{x}$, $\delta _{y}$ and $\delta_{xy}$, 
and treat these parameters as independent random variables.
Finally, we must take into account the laser noise $\delta _{l}$, which is 
rooted in instabilities in the laser cavity (e.g. mode hopping and mode
competition). The laser noise is common to both x and y-polarized light. 
The root mean square values of the fluctuation parameters are defined as 
\begin{equation}
\Delta _{\alpha}=\sqrt{<\delta_{\alpha} ^{2}>}\, , \,\,\, \alpha\,=\, x,\,xy,\,y,\,l \,\,\, .
\end{equation}
For the rest of the paper we will assume that all the signals and fluctuations are 
real-valued. The intensity reaching the individual detectors in Fig. \ref{f2} 
can be given as 
\begin{equation}
I_{1,2}=|\frac{E_{0}}{\sqrt{2}}(1+\delta_{l}) \left[ \gamma E_{x}
(1+\delta _{x}) \pm E_{y} (1+\delta _{y}) \pm 
E_{x}\delta_{xy} \right] |^{2}\,\,\,  ,
\end{equation}
where $+$ and $-$ corresponds to detector 1 and 2, respectively.  
To first order this expression becomes  
\begin{equation}
I_{1} =\frac{1}{2} I_{0}\left[ (\gamma E_{x} + 
E_{y})^{2} + (\gamma E_{x} + E_{y})[E_{x}\gamma
(\delta_{l} +\delta_{x}) +E_{x} \delta_{xy} +E_{y}
(\delta_{l} +\delta_{y})]\right] \,\,\,  ,
\end{equation}
\begin{equation}
I_{2} =\frac{1}{2}I_{0}\left[ (\gamma E_{x} - 
E_{y})^{2} + (\gamma E_{x} - E_{y})[E_{x}\gamma 
(\delta_{l} +\delta_{x}) -E_{x} \delta_{xy} - E_{y}
(\delta_{l} +\delta_{y})] \right]\,\,\,  .
\end{equation}
The difference signal between the two detectors is therefore given by
\begin{equation}
I_{1}-I_{2} =i_{sig}
\left[1+\left( 2\delta_{l}+\delta_{x}+\delta_{y} +\frac{E_{x}}{E_{y}}
\delta_{xy} \right) \right]\,\,\,  ,
\end{equation}
where the useful signal is
\begin{equation}
i_{sig}=2\gamma I_{0} E_{x}E_{y} \,\,\,  .
\end{equation}
The noise due to variations in the reflectivities becomes
\begin{equation}
i_{rms}^{2} =(i_{sig})^{2} \left( 4\Delta _{l} ^{2} + \Delta
_{x} ^{2} + \Delta _{y} ^{2} + (\frac{E_{x}}{E_{y}})^{2}\Delta _{xy} ^{2}
\right)  \,\,\,  .
\end{equation}
Two other inavoidable noise sources are the shot noise and the thermal noise.
The shot noise is expressed by 
\begin{equation}
i_{shot}^{2} = 2e\left| \frac{E_{0}}{\sqrt{2}} \right| ^{2} (\gamma ^{2} E_{x}^{2}
+ E_{y}^{2}) B  = eI_{0} (\gamma ^{2} E_{x}^{2} + E_{y}^{2}) B \,\,\,  ,
\end{equation}
where $B$ is the bandwidth of electronic system. 
The thermal noise is 
\begin{equation}
i_{th}^{2} =  \frac{4kTB}{R} \,\,\,  ,
\end{equation}
where $k$ is the Boltzmann constant, $T$ is the temperature and $R$ is the load
resistance of the differential amplifier. 

In total the noise from the differential system is given by
\begin{equation}
i_{n}^{2} = i_{rms}^{2} + 2i_{shot}^{2} + 2i_{th}^{2}\,\,\,  .
\label{noise}
\end{equation}
Since there are two detectors in the system, the thermal and the 
shot noise from each of them add together.

The signal to noise ratio (SNR), is defined as
\begin{equation}
SNR=\frac{i_{sig}}{i_{n}}
\label{SNR}
\end{equation}
It should be pointed out that the SNR considered here does not take into account
diffraction or interbit crosstalk, and should therefore be regarded as an
upper estimate. 

To illustrate the behaviour of the SNR in the current model, Fig. \ref{f7} shows 
the SNR  as a function of bit size for 3 different flying heights. 
The parameters used here are $I_{0}=1$ mW, $B=10$ MHz, $R=100$ $k\Omega $, 
$t=1$ $\mu m$, $\gamma =0.5$, $\Delta _{x}=\Delta _{y}=0$ and 
$\Delta _{l}=\Delta _{xy}= 10^{-5}$ in addition to indicator parameters 
adopted earlier. The figure suggests that it is necessary to keep the flying
height less than 100 nm in order to obtain a SNR of 30 for bit sizes around 
200 nm. For bit sizes of 1 $\mu m$, the SNR can be as large as 500.

Fig. \ref{f8} shows the SNR as a function of bit size for different
y-polarized noise components $\Delta _{y}$ when $t=1$ $\mu m$ and d=100 nm. 
The main effect of $\Delta _{y}$
seems to be a reduction of the SNR for larger bit sizes, which is due to the
fact that the noise increases with increasing magnetooptic signal. 

Fig. \ref{f9} shows the SNR as a function of flying height for changing 
$\Delta _{y}$ when $t=1$ $\mu m$, $b=0.5$ $\mu m$. The curves indicate that 
$\Delta _{y}$ is very important for small flying heights, and should be kept 
less than $10^{-2}$ to avoid significant reduction of the SNR. 

\section{Resolution}
A critical parameter is the optical resolution of the system. In order to
estimate the resolution, we adopt the method of Ichimura et
al.\cite{Ichimura}. Since the off-axis reflection coefficients are very small,
we neglect their influence on the focusing properties of the system.  
We also set the refractive index of the SIL to n=2, since this is close to the refractive index 
of GGG, and the semiconverging (focusing) angle to $\alpha =60$ degrees. 
This gives NA=1.73, which is within reach of the current technology. 
We see from Fig. \ref{f10} that it is possible to obtain full-width-half-maximum 
resolution of $\sim 200$ nm, although the width is slightly different in the 
x and y-directions. It is important to note that our system is, unlike most
other SIL-systems, not based on evanescent waves, and does not suffer from the
stabillity problems possessed by such systems. Finally it is also useful to 
note that the resolution (and focal depth) may be improved by introducing an 
appropriate phase-apodizer\cite{Helseth1}.

\section{Extension to several channels}
The system discussed above was primarily designed for detection of only one
channel/track. To extend this to $N$ channels, i.e. 
to detect of $N$ magnetic data tracks simultaneously, it is 
necessary to have an optical readout system with larger field of 
view. One method is here to include the indicator films 
in a commerically available magnetooptical disk drive.  A typical lens used in current optical disk drives 
is a molded glass aspheric singlet, designed for operation at single wavelength. It is 
typically 5 mm in diameter, weighs less than 0.1 g, and has a focal length of 
approximately 4 mm\cite{Sasian}. Its NA is 0.5-0.6, and the field of view 
covers a circle with diameter around 100 $\mu m$. To obtain a thin elliptical illumination beam on the 
indicator, a cylindrical lens must be inserted in front of the 
laser. Maeda and Koyonagi found that for an ordinary optical 
disk drive this has negligible influence on the resolution if the NA is 
around 0.5\cite{Maeda}. Then, if the magnetic tracks each has a width of 5 $\mu m$, it 
should be possible to read out 20 tracks simultaneously, each 
with a bit size down to 0.5 $\mu m$. 

Another method for increasing the field of view is to use a cylindrical 
SIL in combination with an imaging lens\cite{Il'yashenko}. A cylindrical 
SIL will give approximately the same resolution as the spherical SIL. 
Then, by using a suitable cylindrical or biconic lens to image the tracks onto
the pixels, a large field of view and a high optical resolution can exist 
simultaneosuly.

\section{Conclusion}
The detection of bits in magnetic media using magnetooptical indicators 
with in-plane magnetization has been investigated. A simple model for the 
magnetooptical detection system was presented. Using this model we have
calculated the Faraday rotation and signal to noise ratio, assuming that the
signal comes from one bit only, and that the system has negligible absorption.
The signal to noise ratio was found to vary strongly with the bit size, the 
polarization noise as well as the distance between the magnetic carrier and the indicator. In
particular, our model predicts that a signal to noise ratio of more 
than 30 is possible for bit sizes of 200 nm at flying heights of less than 100
nm. We also estimate the resolution and the focal depth of the system. It is 
shown that an optical resolution of $\sim 200$ nm is possible using a solid 
immersion lens. 

\acknowledgements
The authors acknowledge helpful discussions with A. Kozlov. This work was 
financially supported by the Norwegian Research Council and Tandberg Data ASA.

\begin{figure}
\caption{Setup for magnetooptic imaging using an indicator film with
in-plane magnetization.
\label{f1}}
\end{figure}

\begin{figure}
\caption{Basic differential detection setup for magnetooptic detection using an 
indicator film with in-plane magnetization.
\label{f2}}
\end{figure}

\begin{figure}
\caption{The magnetooptic conversion process. The dashed lines represent the 
secondary light generated by the magnetooptic conversion. To clarify the
generation of the secondary light, we have given the incoming light a small
angle from the normal, although the analysis in the text assumes normal incident. 
Note that secondary light is generated both on the way down and also on the way
back into the optical system. Thus, the Faraday rotation is ideally doubled.  
\label{f3}}
\end{figure}

\begin{figure}
\caption{Faraday rotation at wavelength $\lambda=633$ nm as a function of magnetic field for 
two different garnet films of composition 1) $Lu_{2.2}Bi_{0.8}Fe_{3.8}Ga_{1.2}O_{12}$ 
and 2) $Lu_{1.7}Y_{0.7}Bi_{0.6}Fe_{3.8}Ga_{1.2}O_{12}$. 
\label{f4}}
\end{figure}

\begin{figure}
\caption{Calculated Faraday rotation as a function of bit size for $\lambda=633$ $nm$ 
when $t=0.5$ $\mu m$ (solid line) and 
$t=5$ $\mu m$ (dash-dotted line). The curves were obtained from 
eq. (\ref{magref}) and eq. (\ref{far}) with  $V=1$ $\mu m ^{-1}$, 
$H_{a} = 60$ $kA/m$ and $M = 100$ $kA/m$.
\label{f5}}
\end{figure}

\begin{figure}
\caption{Calculated Faraday rotation as a function of bit size  
when $d=50$ nm (solid line), $d=200$ nm (dashed line) and $d=500$ nm
(dash-dotted line). The curves were obtained from 
eq. (\ref{magref}) and eq. (\ref{far}) with  $V=1$ $\mu m ^{-1}$, 
$H_{a} = 60$ $kA/m$ and $M = 100$ $kA/m$. 
\label{f6}}
\end{figure}

\begin{figure}
\caption{SNR as a function of bit size 
when $d=500$ nm (solid line), $d=200$ nm (dashed line) and $d=50$ nm
(dash-dotted line). The curves were obtained from 
eq. (\ref{SNR}) with  $V=1$ $\mu m ^{-1}$, 
$H_{a} = 60$ $kA/m$ and $M = 100$ $kA/m$.
\label{f7}}
\end{figure}

\begin{figure}
\caption{SNR as a function of bit size 
when $\Delta _{y}=0$ (solid line), $\Delta _{y}=0.001$ (dashed line) and 
$\Delta _{y}=0.01$ (dash-dotted line). The curves were obtained from 
eq. (\ref{SNR}) with  $V=1$ $\mu m ^{-1}$, 
$H_{a} = 60$ $kA/m$ and $M = 100$ $kA/m$.
\label{f8}}
\end{figure}

\begin{figure}
\caption{SNR as a function of flying height  
when $\Delta _{y}=0$ (solid line) and $\Delta _{y}=0.01$ (dashed line).  The curves were obtained from 
eq. (\ref{SNR}) with  $V=1$ $\mu m ^{-1}$, 
$H_{a} = 60$ $kA/m$ and $M = 100$ $kA/m$.
\label{f9}}
\end{figure}

\begin{figure}
\caption{The normalized time-averaged electric energy density. The solid line
shows the profile of the energy density in the y-direction, whereas the dashed
line corresponds to the x-direction. 
\label{f10}}
\end{figure}

\newpage
\centerline{\includegraphics[width=14cm]{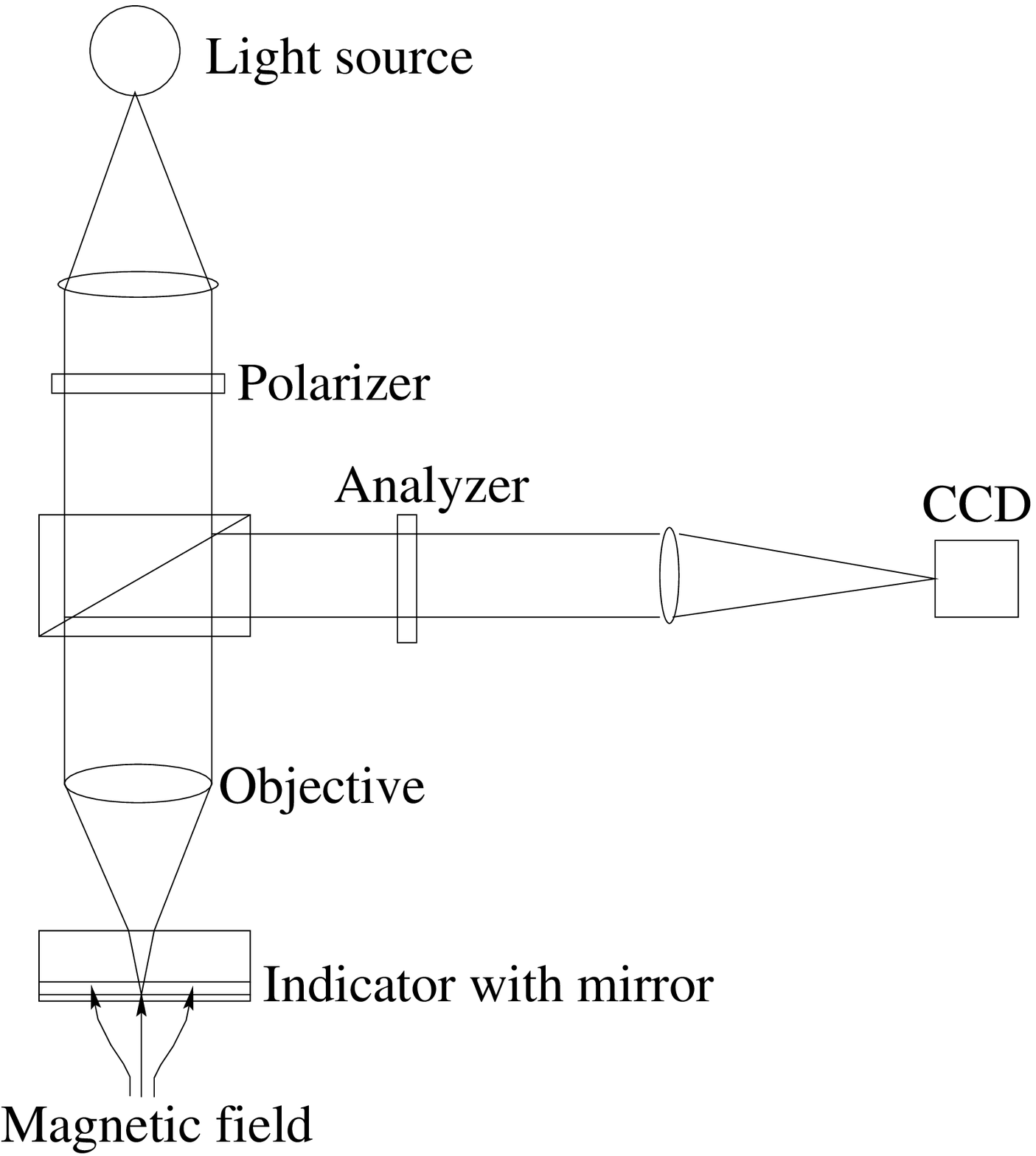}}
\vspace{2cm}
\centerline{Figure~\ref{f1}}

\newpage
\centerline{\includegraphics[width=14cm]{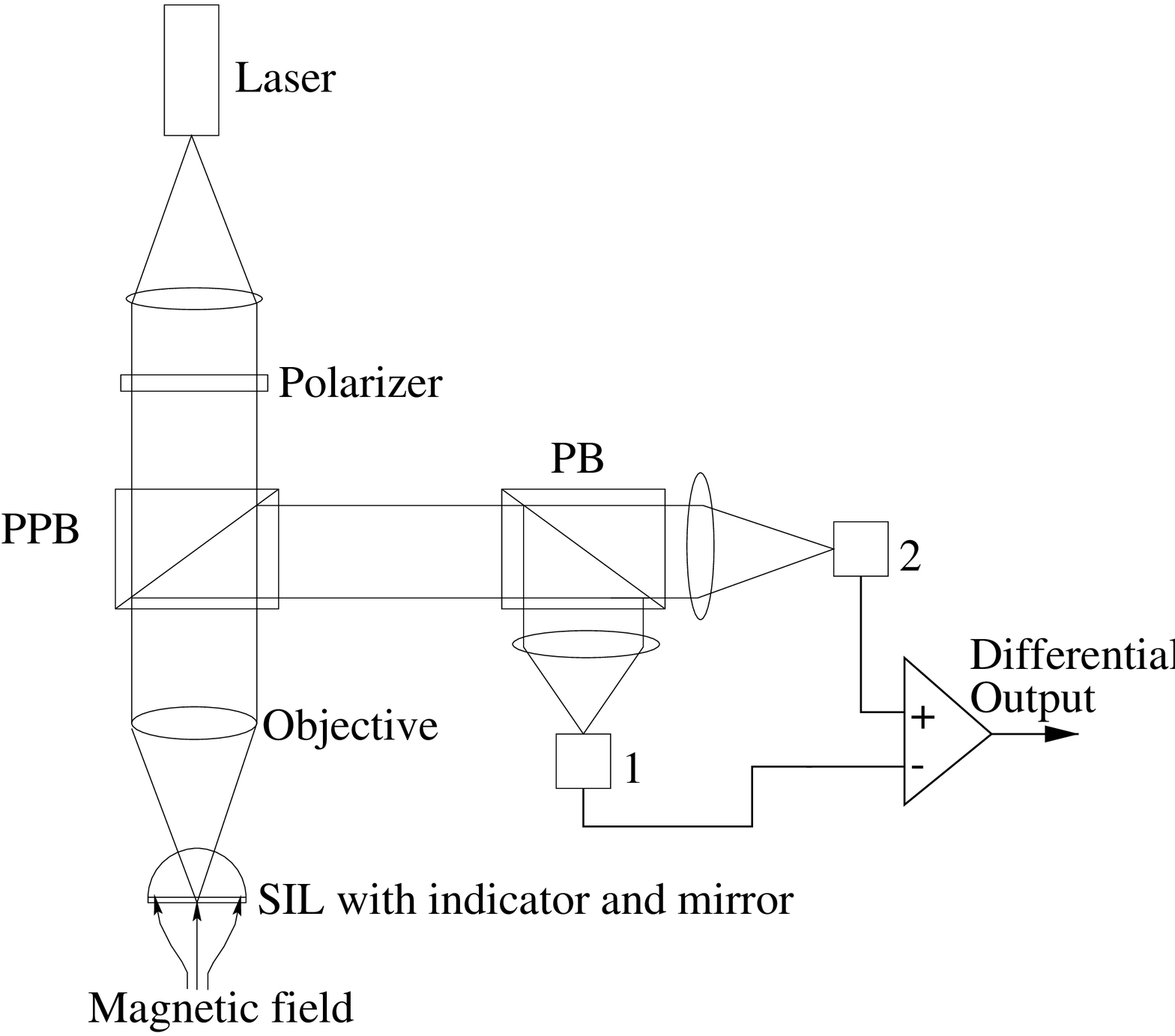}}
\vspace{2cm}
\centerline{Figure~\ref{f2}}

\newpage
\centerline{\includegraphics[width=14cm]{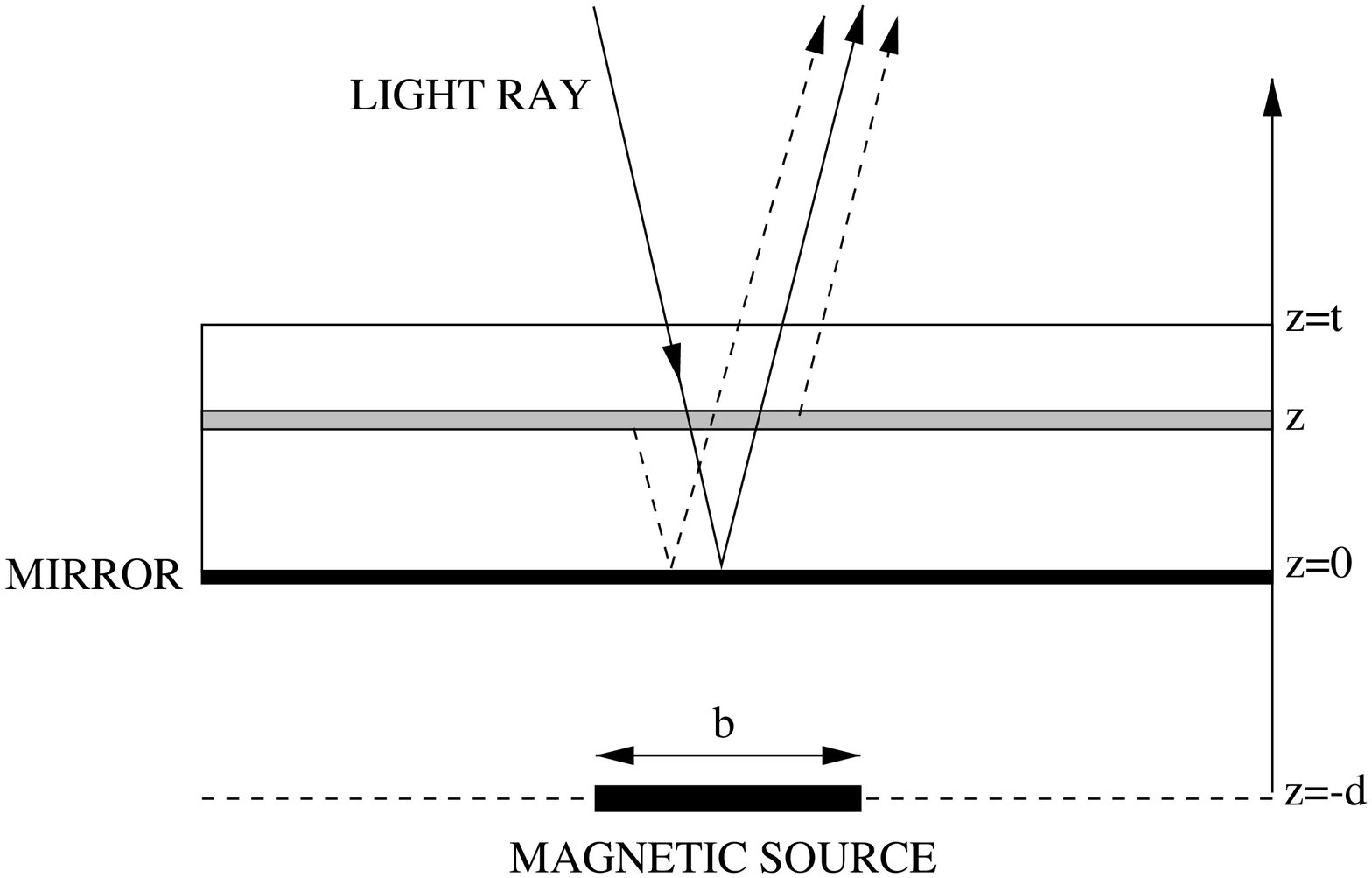}}
\vspace{2cm}
\centerline{Figure~\ref{f3}}

\newpage
\centerline{\includegraphics[width=14cm]{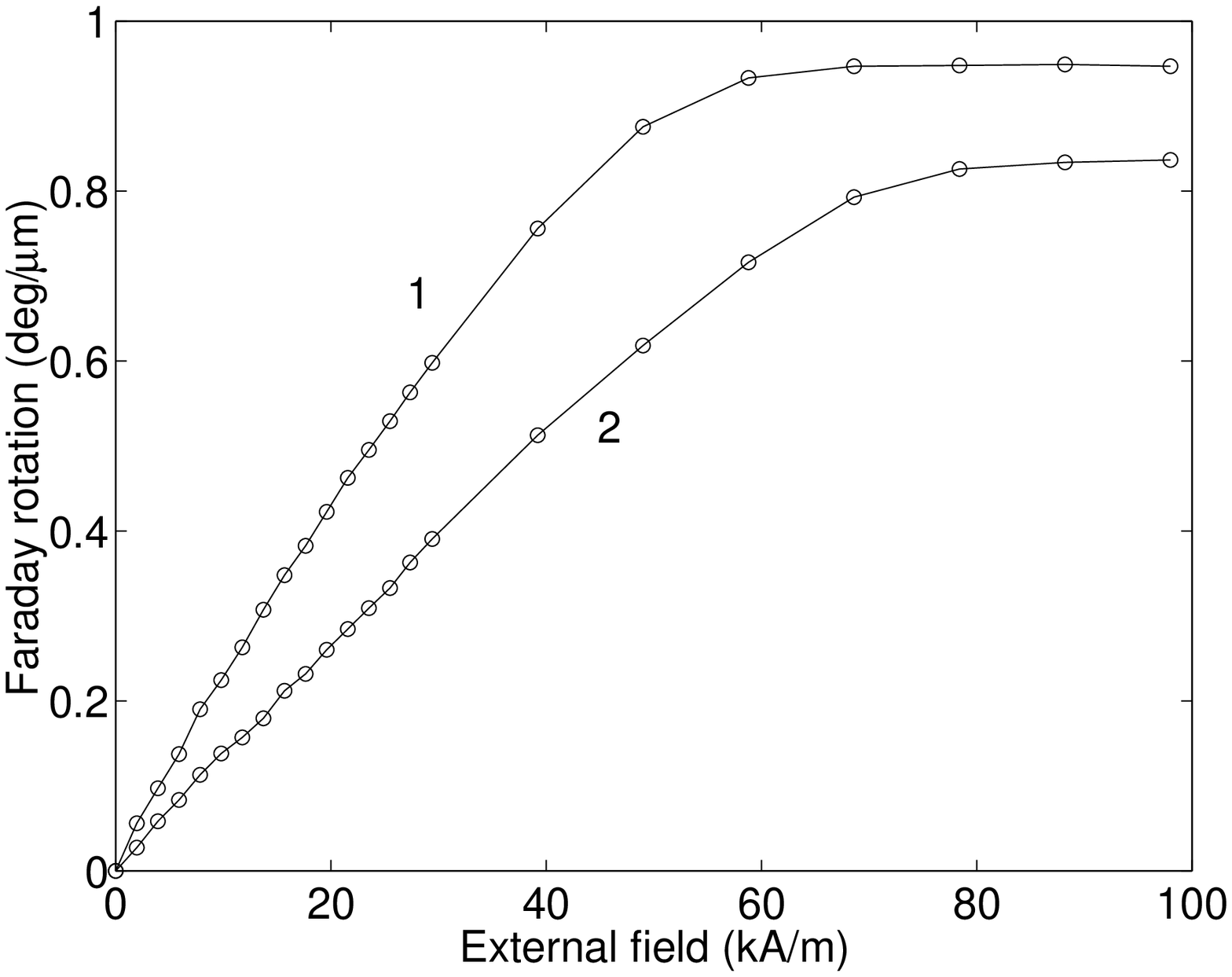}}
\vspace{2cm}
\centerline{Figure~\ref{f4}}

\newpage
\centerline{\includegraphics[width=14cm]{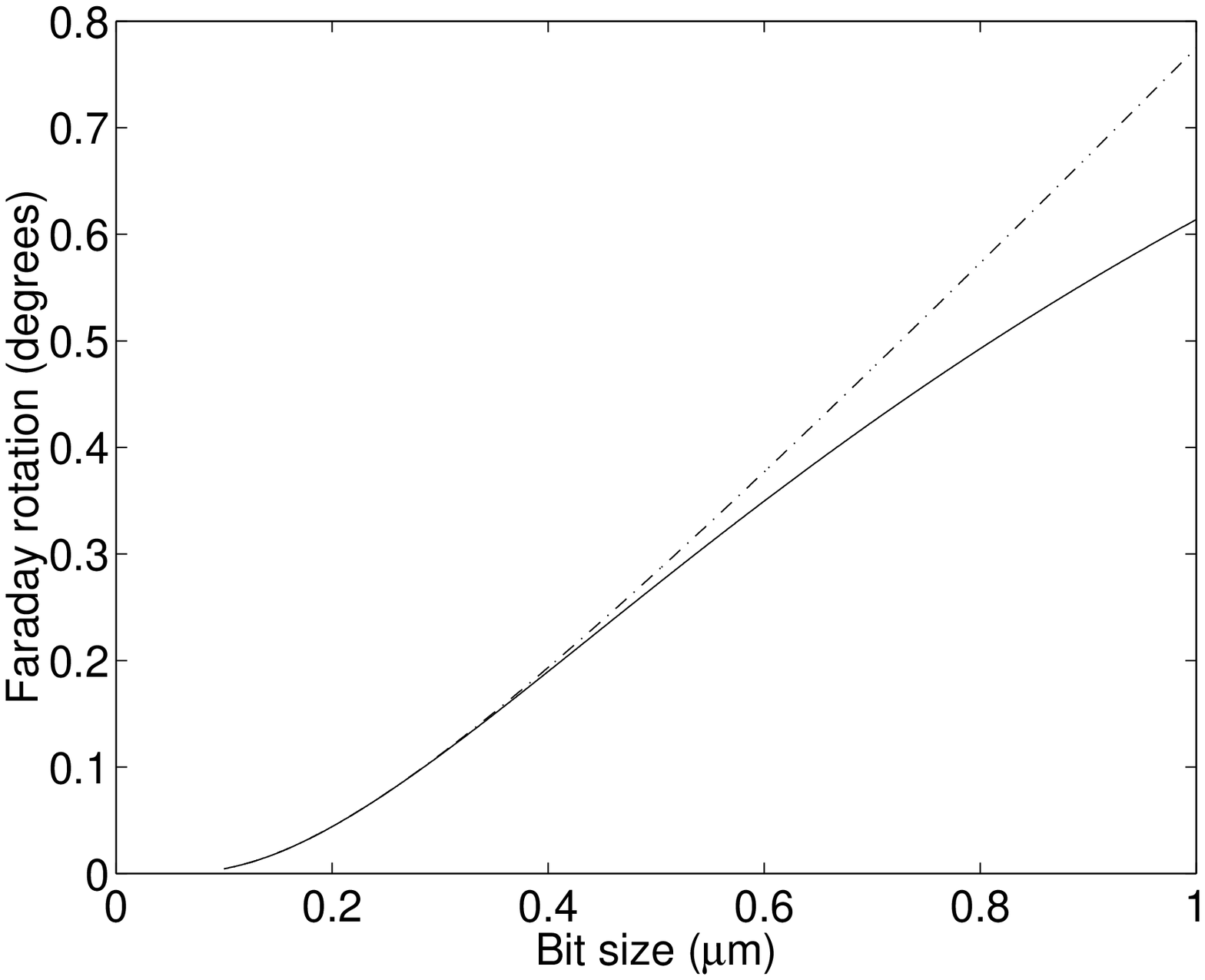}}
\vspace{2cm}
\centerline{Figure~\ref{f5}}

\newpage
\centerline{\includegraphics[width=14cm]{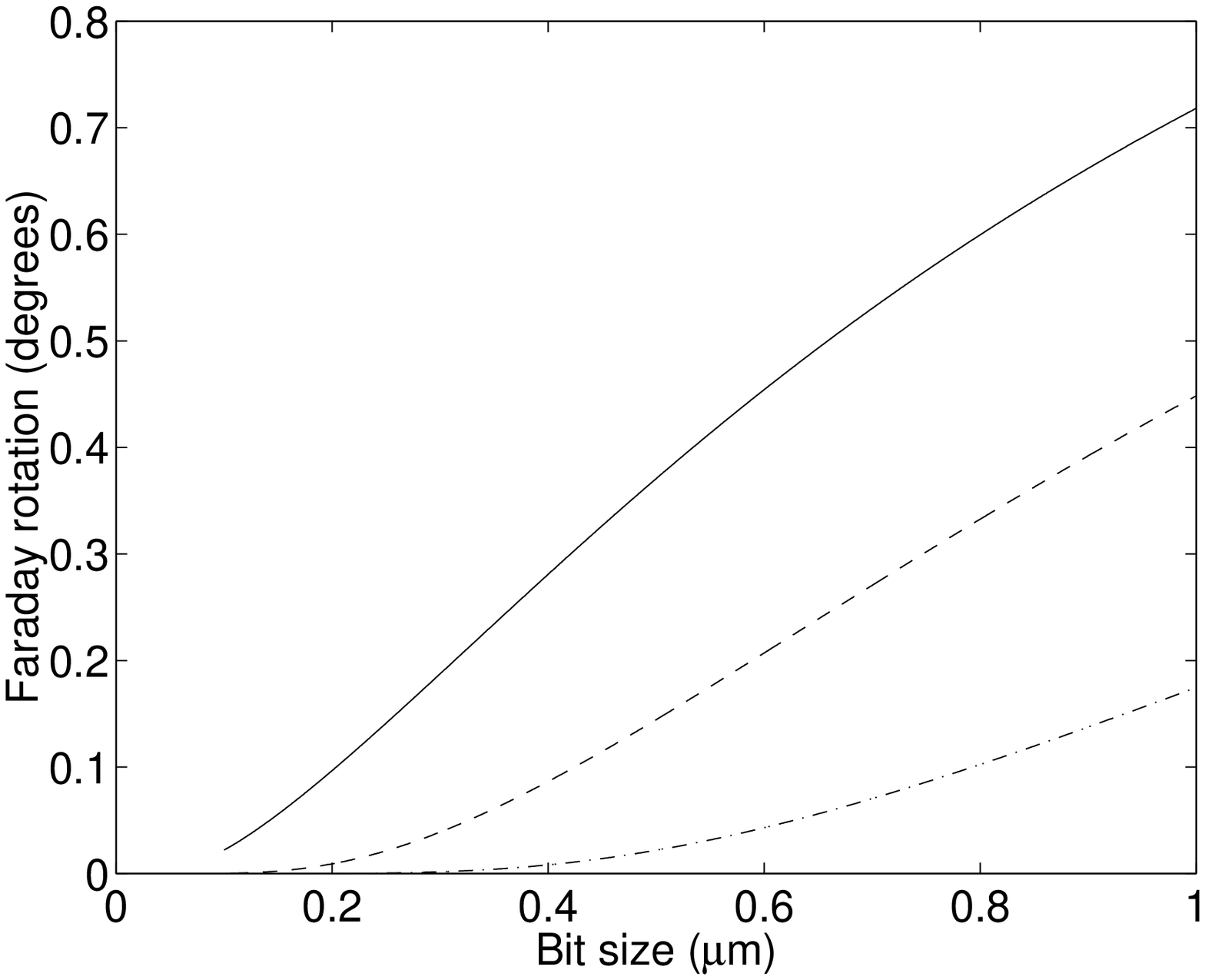}}
\vspace{2cm}
\centerline{Figure~\ref{f6}}

\newpage
\centerline{\includegraphics[width=14cm]{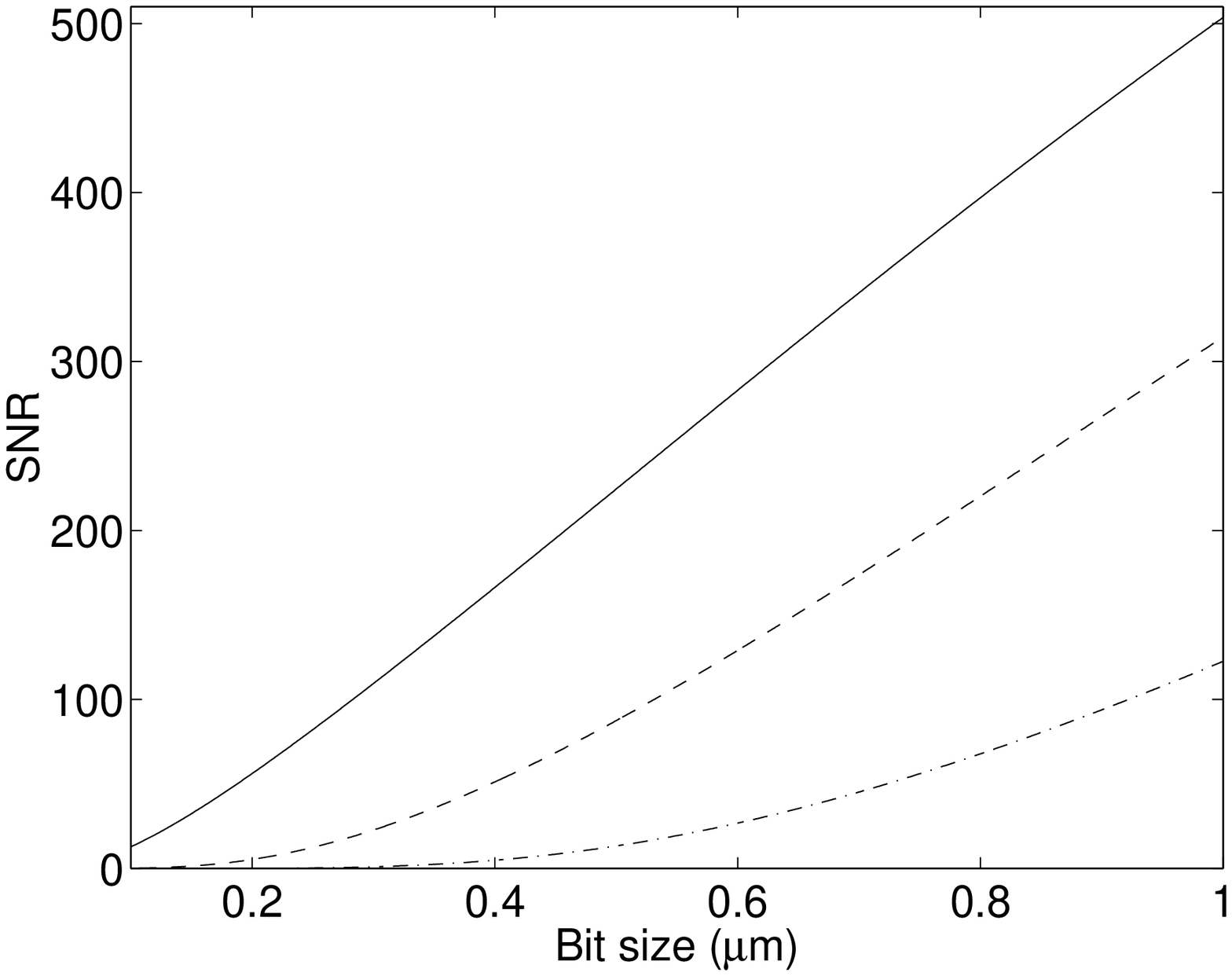}}
\vspace{2cm}
\centerline{Figure~\ref{f7}}

\newpage
\centerline{\includegraphics[width=14cm]{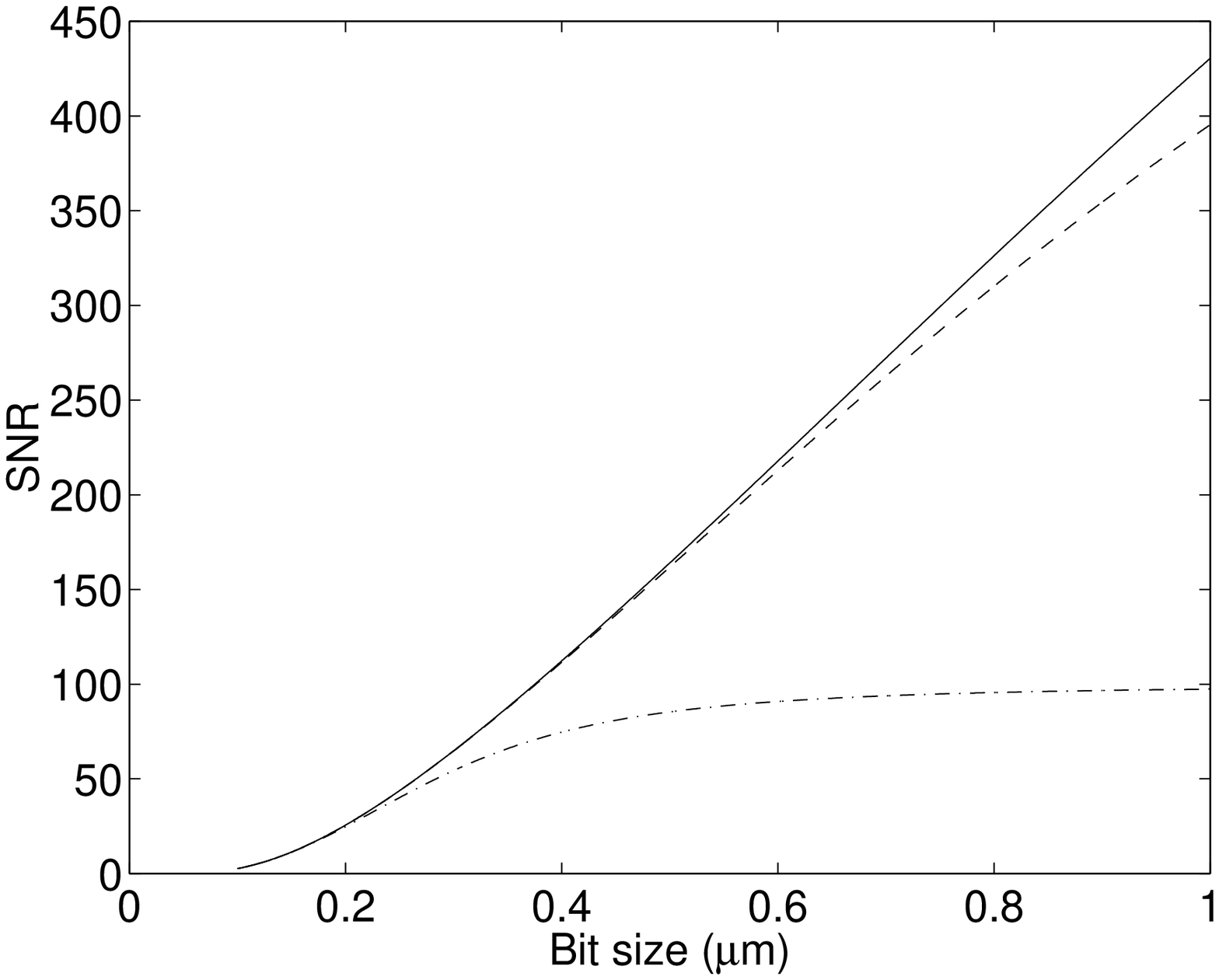}}
\vspace{2cm}
\centerline{Figure~\ref{f8}}

\newpage
\centerline{\includegraphics[width=14cm]{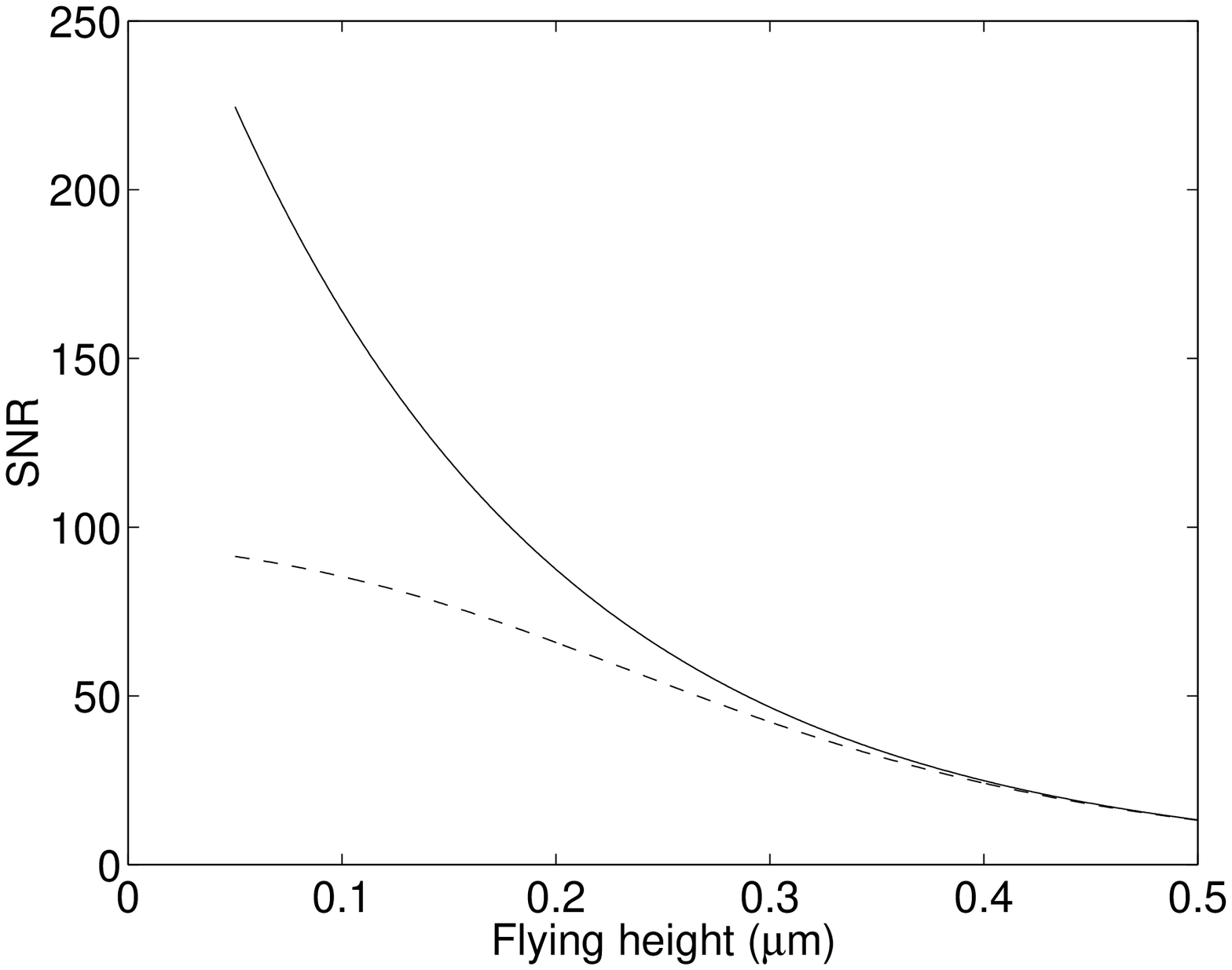}}
\vspace{2cm}
\centerline{Figure~\ref{f9}}

\newpage
\centerline{\includegraphics[width=14cm]{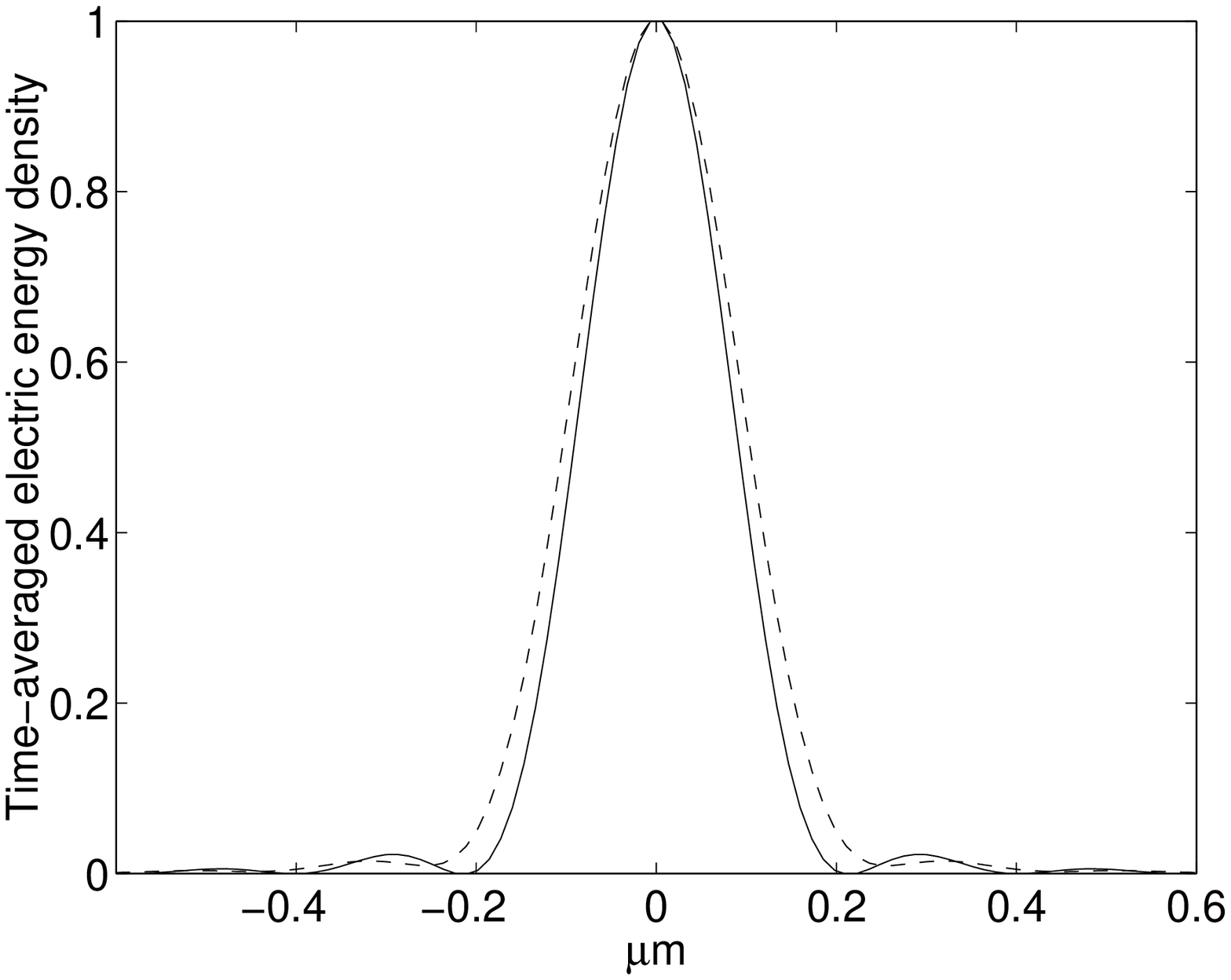}}
\vspace{2cm}
\centerline{Figure~\ref{f10}}

\end{document}